\documentclass[aps,prl,twocolumn,showpacs,preprintnumbers,amsmath,superscriptaddress,amssymb,floats,nofootinbib]{revtex4}

\usepackage{graphicx}
\usepackage{dcolumn}

\begin{document}

\def\Journal#1#2#3#4{{#1} {\bf #2}, #3 (#4)}

\def\NCA{\rm Nuovo Cimento}
\def\NIM{\rm Nucl. Instrum. Methods}
\def\NIMA{{\rm Nucl. Instrum. Methods} A}
\def\NPB{{\rm Nucl. Phys.} B}
\def\PLB{{\rm Phys. Lett.}  B}
\def\PRL{\rm Phys. Rev. Lett.}
\def\PRD{{\rm Phys. Rev.} D}
\def\PRC{{\rm Phys. Rev.} C}
\def\ZPC{{\rm Z. Phys.} C}	
\def\JPG{{\rm J. Phys.} G}
\def\st{\scriptstyle}
\def\sst{\scriptscriptstyle}
\def\mco{\multicolumn}
\def\epp{\epsilon^{\prime}}
\def\vep{\varepsilon}
\def\ra{\rightarrow}
\def\ppg{\pi^+\pi^-\gamma}
\def\vp{{\bf p}}
\def\ko{K^0}
\def\kb{\bar{K^0}}
\def\al{\alpha}
\def\ab{\bar{\alpha}}
\def\be{\begin{equation}}
\def\ee{\end{equation}}
\def\bea{\begin{eqnarray}}
\def\eea{\end{eqnarray}}
\def\CPbar{\hbox{{\rm CP}\hskip-1.80em{/}}}

\title{\large \bf Relaxation of Spin Polarized $^3$He in Mixtures of $^3$He and $^4$He at $\sim$330 mK 
}
\author{Q.~Ye, H.~Gao, W.~Zheng\\
{\it Triangle Universities Nuclear Laboratory and \\
Department of Physics, Duke University, Durham, NC 27708,~USA}\\
D.~Dutta\\
{\it Department of Physics \& Astronomy, Mississippi State University, Mississippi State, MS 39762,~USA}\\
F.~Dubose, R.~Golub, P. Huffman, C.~Swank\\
{\it Department of Physics, North Carolina State University, Raleigh, NC 27695,~USA}\\
 E.~Korobkina\\
{\it Department of Nuclear Engineering, North Carolina State University, Raleigh, NC 27695,~USA}\\
}
\begin{abstract}

We report the measurements of depolarization probabilities of polarized $^3$He in a rectangular acrylic cell
with a deuterated tetraphenyl butadiene-doped deuterated polystyrene coating filled with superfluid $^4$He
at $\sim$330 mk with a magnetic holding field of $\sim$7.3 G. We achieve a wall depolarization probability of $\sim1.0\times10^{-7}$. Such a surface will find application in a new experiment searching for the neutron electric dipole moment and other applications.

\end{abstract}


\maketitle

\section{I. Introduction}

Relaxation of spin polarized $^3$He has been studied extensively on different surfaces (mostly in glass containers) at different temperatures~\cite{Schmiedeskamp,Deninger,Piegay,Himbert,Lusher1,Lusher2,lowe,Chapman,Chapman2} in the past several decades. Models have been developed and efforts have been carried out~\cite{Himbert,Lusher1,Lusher2,Babcock,Steenbergen,Newbury,McGregor,Cates,LefevreBrossel} to understand and suppress the sources of depolarization for different applications~\cite{GentileChen,Stoner,Greene,Stirling,npdg,bear,edm}.
Among these applications, the search for violations of fundamental symmetries~\cite{npdg,bear,edm} will have a deep impact on our understanding of the universe.
For example, in a new search for the neutron Electric Dipole Moment (nEDM)~\cite{edm},
polarized $^3$He dissolved in superfluid $^4$He will be exposed to unique experimental conditions, which have never been studied before and the success of the experiment requires a $^3$He longitudinal relaxation time in excess of tens of times the neutron life time.
The search for the neutron EDM is a direct search of the time reversal symmetry (T) violation. It provides a unique way of searching for charge conjugation and parity symmetry (CP) violation due to the CPT invariance.

The overall experimental strategy for the new nEDM experiment is to form a three-component fluid of polarized ultra-cold neutrons (UCN) and polarized $^{3}$He atoms dissolved in a bath of superfluid $^{4}$He below 500 mK
with a $^3$He to $^4$He ratio $\sim 1:10^{10}$~\cite{edm}. 
The nEDM manifests itself as a difference in the neutron precession frequencies when a strong electric field parallel to an external magnetic field is reversed. Because the magnitude of the precession frequency shift due to the interaction of the nEDM and the electric field is extremely small, the new experiment proposed will measure the UCN precession frequency relative to that of the $^3$He nucleus~\cite{edm} using the spin-dependent nuclear absorption process:
\begin{equation}
\vec{n}+\vec{^{3}He}\rightarrow p+t+764~keV
\label{eq1}
\end{equation} 
When placed in an external magnetic field, both the neutron and $^3$He magnetic dipoles will precess in the plane perpendicular to the $B_0$ field. The $^3$He nucleus' EDM is highly suppressed due to ``Schiff Shielding''~\cite{schiff}.
The neutron absorption rate can be measured by monitoring the Extreme Ultra-Violet (EUV) scintillation light generated in superfluid $^4$He from the reaction products in (\ref{eq1}).
The polarized $^3$He nucleus also acts as a co-magnetometer
and its signal is monitored using Superconducting QUantum Interference Devices (SQUIDs).
The knowledge of the $B_0$ field environment of the trapped neutrons is important in the analysis of the systematic uncertainties of the experiment.

The neutron storage cell will be made of acrylic with deuterated TetraPhenyl Butadiene-doped deuterated PolyStyrene (dTPB-dPS) coating on the inner wall. The dTPB-dPS coating is chosen to minimize the UCN absorption on the wall, which also acts as a wavelength shifting material for the EUV light.
Therefore, understanding the relaxation mechanism of polarized $^3$He
in the measurement cell under the unique nEDM experimental conditions and suppressing the depolarization effect is crucial to the experiment.

In our previous depolarization study of polarized $^3$He in superfluid $^4$He at 1.9 K~\cite{Ye2008}, we reported a depolarization probability $\sim1.6 \times 10^{-7}$ from a dTPB-dPS coated and cylindrical-shaped acrylic cell. This number may change at lower temperatures. The nEDM experimental cell is of rectangular shape and the experimental temperature is below 500 mK. In this paper we report new results on relaxation times of polarized $^3$He at a temperature down to $\sim$330 mK in a rectangular dTPB-dPS coated acrylic cell filled with superfluid $^4$He.

\section{II. Experimental Apparatus}

\begin{figure}[htbp]
\includegraphics*[width=6cm,height=10cm]{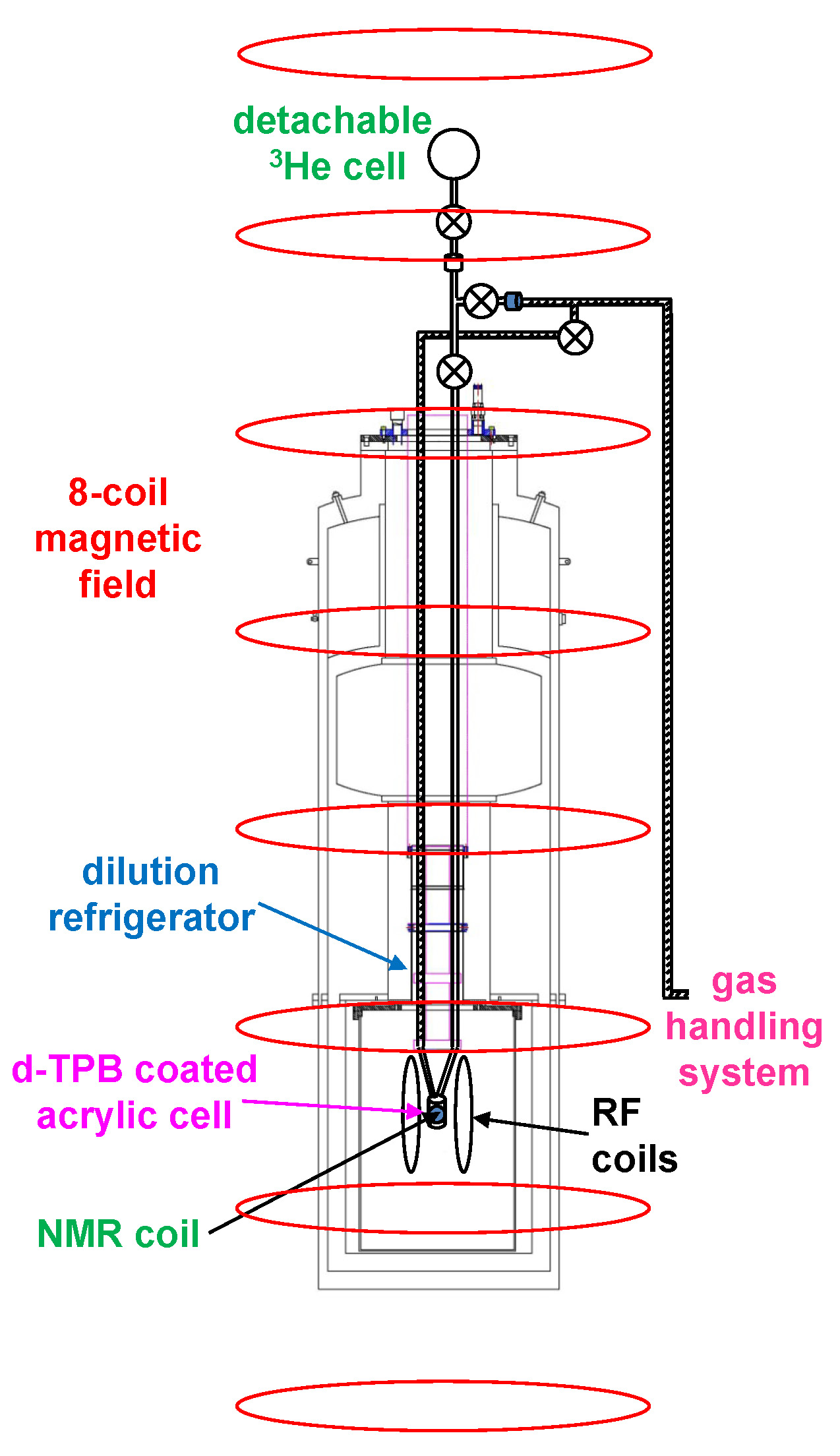}
\caption{(Color online) A schematic of the experimental setup.}
\label{apparatus}
\end{figure}

The schematic of the entire apparatus is shown in Figure~\ref{apparatus}.
Polarized $^3$He atoms are introduced into the bottom cell from a 2 in. diameter detachable pyrex glass cell sitting on top of the dilution refrigerator (DR).
The cell is usually filled with $\sim$1-1.5 atmospheres of $^3$He and $\sim$100 torrs of N$_2$ gas at room temperature. 
The spin-exchange optical pumping (SEOP) technique is used to polarize the $^3$He on the polarizing station located in another building.
The detachable cell is then transferred over to the top of the DR using a battery powered portable magnetic field system.
The bottom acrylic cell is of a rectangular (1''$\times$1''$\times$1.5'') shape with many grooves cut on the outside to house the cooling wires from the mixing chamber of the DR. The cell's inner surface is coated with the dTPB-dPS material and the coating procedure is described in the appendix of~\cite{YePhD2008}. The two cells are connected via a $\sim$86 in. long
glass capillary and separated by a glass valve. A 1/8 in. OD and 1 in. long glass-to-copper seal is used to connect the glass capillary and the acrylic cell to serve as the transition between the two materials with different coefficients of thermal expansion. 
Epoxy Emerson \& Cuming Stycast 1266 is applied to make the vacuum seal between copper and acrylic. A 7.5-liter aluminum volume with a baratron is mounted onto the gas handling system to calibrate the amount of $^4$He condensed into superfluid in the acrylic cell.

A model ``Minikelvin 126-TOF'' dilution refrigerator made by Leiden Cryogenics is used to achieve the desired temperature. 
The DR and its outside dewars are located inside an eight-coil system (33 in. diameter, 16.5 in. separation) providing a uniform magnetic field in the vertical direction. In order to compensate for the edge effect, the current in the outer two coils is larger than that flowing in the middle six coils and the configuration produces $\sim$7.3 G magnetic field at the acrylic cell position. The DR dewar is composed of four layers: the outer vacuum chamber, the liquid N$_2$ layer, the liquid $^4$He main bath and the inner vacuum chamber. The working principle and the operating procedures of the DR can be found in~\cite{Pobell}. 
A 20 cc, gold plated copper buffer volume is attached to the bottom of the mixing chamber, the coldest part of the DR at normal operating conditions. The pre-cooled $^4$He liquid used to fill up the acrylic cell goes into the buffer volume first, then drips into the acrylic cell slowly through a stainless steel capillary. 
Attached to the bottom of the buffer volume is an oxygen free copper cap with twenty gold plated 99.999\% pure copper wires brazed in. The wires are extended onto the outside of the acrylic cell, where wire housing grooves are made and vacuum grease is used to increase the thermal contact between the copper wires and the acrylic. The temperature of the acrylic cell is monitored by calibrated ruthenium oxide sensors placed
on the acrylic cell and the glass-to-copper seal.
During normal operation of the DR, the acrylic cell filled completely with superfluid $^4$He can be cooled down to as low as $\sim$330 mK.

Two sets of nuclear magnetic resonance systems, adiabatic fast passage (AFP) and free induction decay (FID), 
have been used to measure the polarization of $^3$He. The AFP system consists of a pair of 9'' Helmholtz RF coils, mounted below the copper buffer volume to avoid the eddy current heating, and a pick up coil, which is glued onto the side of the acrylic cell. An Apollo LF-1 MRI system from Tecmag is used to carry out the FID measurements at the frequency of 23.6 kHz. The AFP and FID principles can be found in~\cite{Abragam}.

\section{III. Experimental Procedures}

The acrylic cell pieces are made and coated with the dTPB-dPS material. A series of strict leak tests are carried out after the system is assembled.
It then takes about one week to cool down the DR to the operating temperature. Once the acrylic cell has reached $\sim$330 mK, gaseous $^4$He is condensed into superfluid $^4$He and fills up the acrylic cell slowly through the calibrated volume and the DR.
At the same time, the top detachable cell filled with $^3$He is being polarized in the laser lab overnight.
After the polarized detachable cell is brought over onto the top of the DR using the portable magnetic field and the intermediate region is pumped to vacuum, the valve is opened to allow the polarized $^3$He atoms to diffuse to the bottom acrylic cell at low temperatures through the glass capillary. After the system reaches equilibrium in $\sim$6 minutes, which is comparable to the spin diffusion time in our cell geometry~\cite{Opfer}, a series of NMR AFP or FID measurements are carried out to measure the $^3$He longitudinal relaxation time ($T_1$).

The $^3$He detachable cells have a $T_1$ of $\sim$40 hours with a holding magnetic field of 7.3 G and the depolarization during the 10-minute transport process can be neglected. 
A short RF pulse at the Larmor frequency is generated by the FID coil and the $^3$He free precession signal is monitored using the same coil after the pulse. 	
The systematic error of the relaxation time is 
dominated by the uncertainty in the determination of the FID inefficiency, which is derived by 
fitting $n$ consecutive FID measurements to $A_0 (1-loss)^n$, where $loss$ is the FID inefficiency parameter. The relaxation times are obtained from the exponential decay of the FID signal strength as a function of time corrected by the FID inefficiency.
The NMR-AFP system uses an RF amplifier to power the RF coils for the spin flip during the measurement~\cite{Ye2008}. The RF amplifier is left off during the time intervals between AFP measurements to prevent the polarized $^3$He atoms from depolarizing due to the wide band amplified RF noise. The AFP inefficiency and the corresponding relaxation time are determined in similar ways as those of FID measurements.
The measured relaxation times using AFP and FID are consistent with each other within experimental uncertainties.

\begin{figure}[htbp]
\includegraphics*[width=7.5cm]{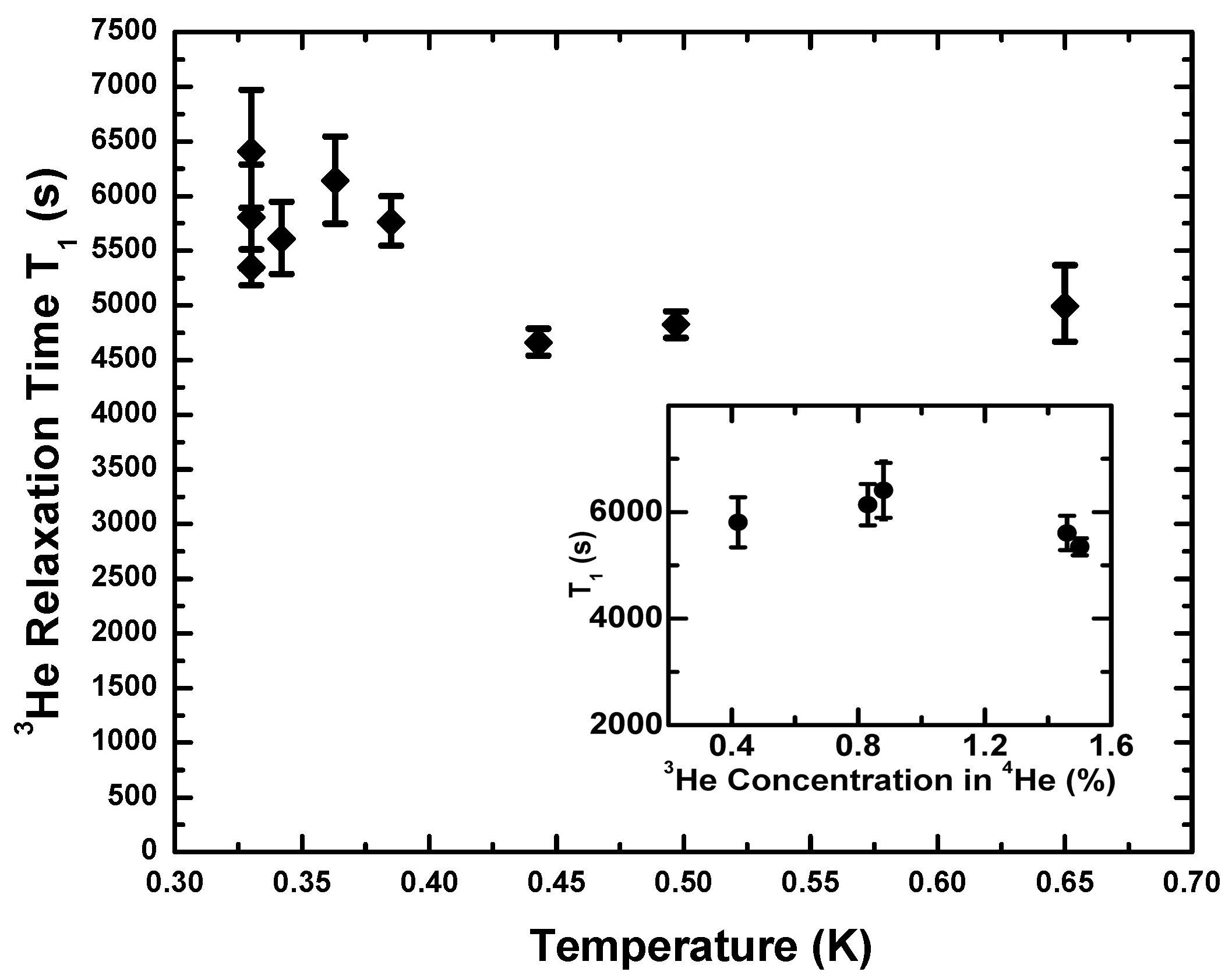}
\caption{$^3$He longitudinal relaxation time versus temperature
in a rectangular dTPB-dPS coated acrylic cell with a holding magnetic field of 7.3 G. The acrylic cell is filled with $\sim$0.89 mole superfluid $^4$He. The error bars are the quadrature sum of the statistical and systematic uncertainties. The figure inset shows the 
$^3$He $T_1$ as a function of the $^3$He molar concentration in superfluid $^4$He $\chi$ at $\sim$330 mK.}
\label{Temp_T1-3HeConcen_T1}
\end{figure}

\section{IV. Results \& Discussion}

In Figure~\ref{Temp_T1-3HeConcen_T1}, $^3$He longitudinal relaxation times are plotted as a function of temperature at 7.3 G. The inset of the figure shows the measured $^3$He $T_1$ versus the $^3$He molar concentrations in superfluid $^4$He, which indicates that $^3$He $T_1$ does not change much in the $\chi$ ranging from 0.42\% to 1.5\% at $\sim$330 mK.
\begin{figure}[htbp]
\includegraphics[width=7.5cm]{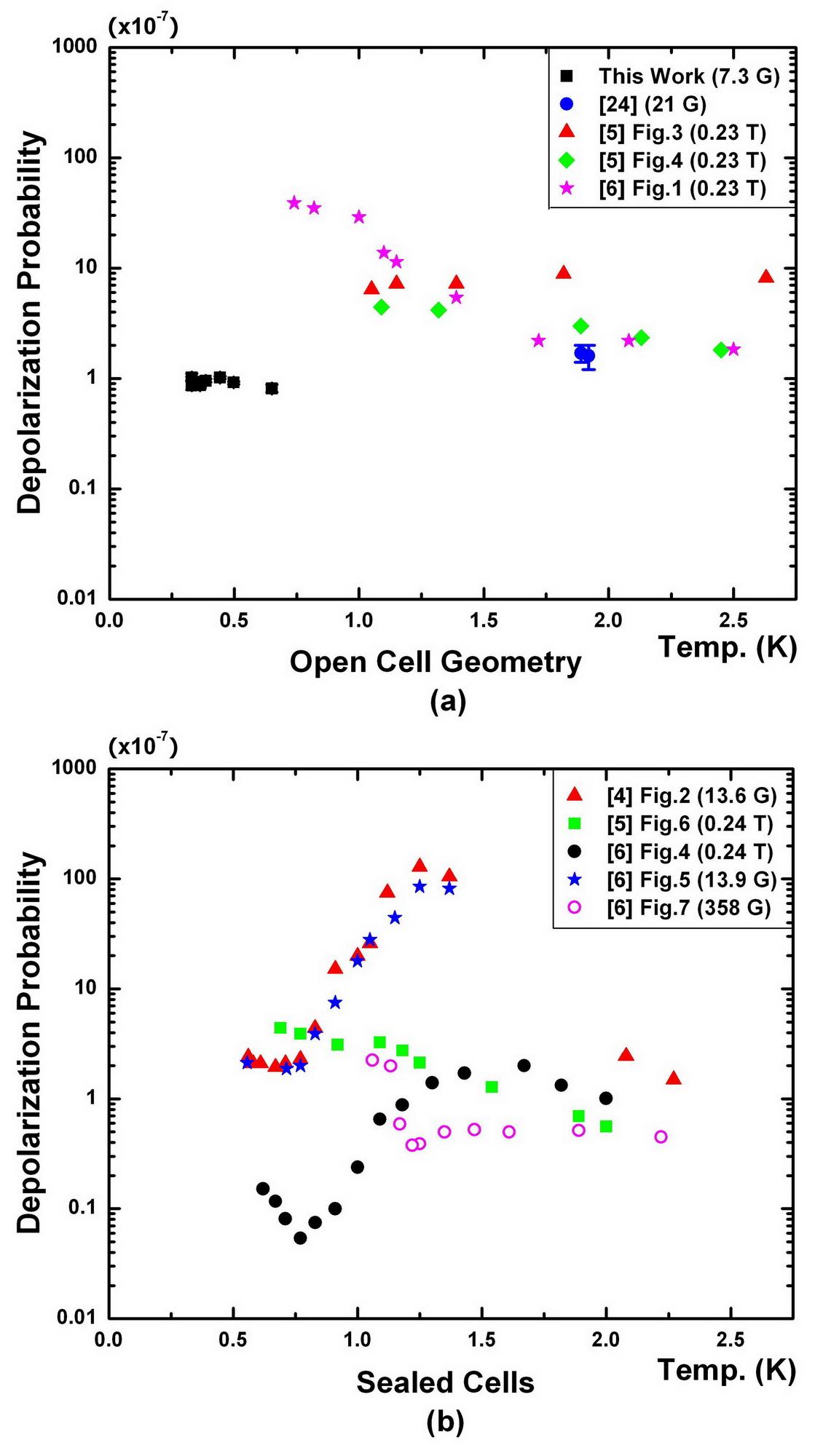}
\caption{(Color online) (a) shows the $^3$He depolarization probabilities versus temperature in this work (black squares) with all $^3$He atoms in bulk superfluid $^4$He. (a) also includes the derived DPs from~\cite{Ye2008} (blue circles), 
~\cite{Lusher1} figure 3 (red triangles), ~\cite{Lusher1} figure 4 (green diamonds) and ~\cite{Lusher2} figure 1 (pink stars). All these data are obtained from open cell systems. (b) lists the derived DPs from sealed cells for comparison, including~\cite{Himbert} figure 2 (red triangles), ~\cite{Lusher1} figure 6 (green squares), ~\cite{Lusher2} figure 4 (black circles), ~\cite{Lusher2} figure 5 (blue stars) and~\cite{Lusher2} figure 7 (pink open circles).
In~\cite{Himbert,Lusher1,Lusher2}, $^3$He atoms are mostly in vapor phase.
}
\label{3HeDPvsT}
\end{figure}
The $^3$He dipole-dipole relaxation time is determined to be higher than 4$\times 10^4$ seconds~\cite{Newbury} with the $^3$He concentration in our measurements. The magnetic field gradient is measured to be $\sim$15 mG/cm, which corresponds to a $^3$He longitudinal relaxation time of $\sim2\times 10^{6}$ seconds~\cite{Cates}. These numbers are much larger than the measured relaxation times and therefore the wall effect is the dominant contribution to the $^3$He relaxation time.
A depolarization probability (DP) can be derived analytically to characterize how ``friendly'' or ``unfriendly'' a surface is to polarized $^3$He.

Consider a sample of 100\% polarized $^3$He atoms in a container of volume $V$ and surface area $S$. The average number of $^3$He atoms colliding with the wall per unit time per unit area is $\frac{1}{4}n\cdot \bar{v}$, where $n$ is the $^3$He number density, $\bar{v}=\sqrt{\frac{8 k_B T}{\pi m^{*}_3}}$ is the mean speed of $^3$He quasiparticles and $m^{*}_3=2.4 m_3$ is the effective mass of $^3$He dissolved in superfluid $^4$He. A polarized $^3$He atom will have a probability DP to depolarize after each collision with the cell surface covered with bulk liquid $^4$He and DP is given by
\begin{equation}
	DP = \frac{4}{\bar{v}\cdot T_1 \cdot (S/V)}
\label{DPT2}
\end{equation}
where $S/V$ is the surface to volume ratio of a particular cell.

Figure~\ref{3HeDPvsT} shows the corresponding wall depolarization probabilities derived from equation (\ref{DPT2}) at different temperatures from our measurements with the dipole-dipole 
depolarization effect discussed earlier taken into consideration. Since our cell is full of superfluid $^4$He during the measurement, 
the surface to volume ratio, $S/V=2.1$, is a constant at different temperatures. Figure~\ref{3HeDPvsT} (a) also includes the data at 1.9 K~\cite{Ye2008}.
The other DPs in the figure are extracted from the data in~\cite{Himbert,Lusher1,Lusher2} by using the $^3$He mean speeds at the corresponding temperatures, the relaxation times and the surface to volume ratios in the open cell geometries and sealed cells.
The polarized $^3$He in our cell is in the bulk liquid $^4$He with layers of superfluid $^4$He film on the wall surfaces, while in~\cite{Himbert,Lusher1,Lusher2}, $^3$He atoms are mostly in vapor phase.
Our data show that the DPs from the dTPB-dPS coated acrylic cell surface covered by superfluid $^4$He remain around 1.0$\times10^{-7}$ in the temperature range between 0.33 K and 0.65 K, which is close to the value at 1.9 K~\cite{Ye2008}.
Our experiments probe the DPs down to the lowest temperature of 330 mK and the dTPB-dPS coated acrylic surface outperform the surfaces in open geometry systems (Figure~\ref{3HeDPvsT} (a)), including
bare pyrex glass surfaces (Fig. 3 in~\cite{Lusher1}), glass surfaces with solid molecular hydrogen wall coating (Fig. 4 in~\cite{Lusher1}), and solid H$_2$-coated pyrex surfaces covered by superfluid $^4$He film (Fig. 1 in~\cite{Lusher2}).
In Figure~\ref{3HeDPvsT} (b), DPs from a sealed cell (black circles) reach below $10^{-7}$ from 0.6 K to 1.2 K with a magnetic field of $\sim$0.24 T.
However, the DPs extracted from the same sealed cell become much larger under a magnetic holding field of $\sim$13.9 G (blue stars), which are outperformed by our results with similar magnetic holding fields from an open cell geometry. No magnetic field dependence in the DP is observed in our measurements though our magnetic field range is rather limited. 
The effect due to the background magnetic field gradient in~\cite{Lusher2} may have played a more significant role which resulted in different $^3$He DPs at different holding fields of 0.24 T and 13.9 G.

\section{V. Summary}

We have measured the relaxation time of polarized $^3$He in a dTPB-dPS coated acrylic cell filled with superfluid $^4$He at 330 mK with a magnetic holding field of 7.3 G. The corresponding wall depolarization probability is on the order of 1.0$\times10^{-7}$ for polarized $^3$He.
The nEDM experimental cell will have a $S/V$ ratio of $\sim$0.5 cm$^{-1}$~\cite{edm}, so the extrapolated relaxation time of polarized $^3$He in the nEDM cell geometry is $\sim2.5\times 10^{4}$ s at 330 mK.
This long relaxation time has met the stringent requirement of the experiment~\cite{edm}, and also makes dTPB-dPS coated acrylic surface potentially important for other applications.

\section{Acknowledgment}

We thank C. Arnold, V. Cianciolo, T. Clegg, D. Haase and M. Hayden for helpful discussions, M. Cooper for support, R. Lu, X. Zhu, X. Zong for help with the experiment, J. Rishel for making the glassware and J. Addison, B. Carlin, J. Dunham, R. O'Quinn, P. Mulkey and C. Westerfeld for the technical support. We also thank Hahn Meitner Institute of Berlin for the loan of the DR. This work is supported by the School of Arts and Science of the Duke University, the U.S. Department of Energy under contract number DE-FG02-03ER41231 and the Los Alamos National Laboratory.

\end{document}